\documentclass[journal,twoside,confmode]{ieeecolor}
\usepackage{tmi}
\usepackage{cite}
\usepackage{hyperref}
\usepackage{amsmath,amssymb,amsfonts}
\usepackage{algorithmic}
\usepackage{gensymb}
\usepackage{url}
\usepackage{afterpage}

\usepackage{graphicx}
\usepackage[ruled,linesnumbered]{algorithm2e}
\usepackage{bm}
\usepackage{textcomp}
\usepackage{graphicx}
\usepackage{caption}
\usepackage{array}
\usepackage{subcaption}
\usepackage{multirow}
\def\BibTeX{{\rm B\kern-.05em{\sc i\kern-.025em b}\kern-.08em
    T\kern-.1667em\lower.7ex\hbox{E}\kern-.125emX}}
\begin{document}
\title{Physics Driven Domain Specific Transporter Framework with Attention Mechanism for Ultrasound Imaging}
\author{Arpan Tripathi, Abhilash Rakkunedeth, Mahesh Raveendranatha Panicker, Jack Zhang, Naveenjyote Boora, Jessica Knight, Jacob Jaremko, Yale Tung Chen, Kiran Vishnu Narayan, Kesavadas C

\thanks{This work was supported in part by the Department of Science and Technology - Science and Engineering Research Board (DST SERB (CVD/2020/000221)) and experiments enabled by Compute Canada Cluster}
\thanks{Arpan Tripathi and Mahesh Raveendranatha Panicker are with Indian Institute of Technology Palakkad, Abhilash Rakkunedeth (e-mail:hareendr@ualberta.ca), Jack Zhang, Naveenjyote Boora, Jessica Knight and Jacob Jaremko are with Department of Radiology, University of Alberta, Yale Tung Chen is with Hospital Universitario Puerta de Hierro Spain, Kiran Vishnu Narayan is with Government Medical College Thiruvananthapuram and Kesavadas C is with Sree Chitra Tirunal Institute for Medical Sciences \& Technology Thiruvananthapuram.}
}
\maketitle
\begin{abstract}
Most applications of deep learning techniques in medical imaging are supervised and require a large number of labeled data which is expensive and requires many hours of careful annotation by experts. In this paper, we propose an unsupervised, physics driven domain specific transporter framework with an attention mechanism to identify relevant landmarks with applications in ultrasound imaging. The proposed framework identifies key points that provide a concise geometric representation highlighting regions with high structural variation in ultrasound videos. We incorporate physics driven domain specific information as a feature probability map and use the radon transform to highlight features in specific orientations.The proposed framework has been trained on 130 Lung ultrasound (LUS) videos and 113 Wrist ultrasound (WUS) videos and validated on 100 Lung ultrasound (LUS) videos and 58 Wrist ultrasound (WUS) videos acquired from multiple centers across the globe. Images from both datasets were independently  assessed by experts to identify clinically relevant features such as A-lines, B-lines and pleura from LUS and radial metaphysis, radial epiphysis and carpal bones from WUS videos. The key points detected from both datasets showed high sensitivity (LUS = 99\%, WUS = 74\%) in detecting the  image landmarks identified by experts. Also, on employing for classification of the given lung image into normal and abnormal classes, the proposed approach, even with no prior training, achieved an average accuracy of $97\%$ and an average F1-score of $95\%$ respectively on the task of co-classification with 3 fold cross-validation. With the purely unsupervised nature of the proposed approach, we expect the key point detection approach to  increase the applicability of ultrasound in various examination performed in emergency and point of care. 
\end{abstract}

\begin{IEEEkeywords}
Transporter Neural Network, Unsupervised Learning, Keypoint Detection, Feature Probability Map, Wrist Fracture, Lung Ultrasound
\end{IEEEkeywords}

\section{Introduction}
\label{sec:introduction}
Ultrasound imaging is well suited for emergency care applications as it is easily portable, free from ionizing radiation and inexpensive. It is a viable alternative to x-ray for diagnostic exams performed in emergency care like  fracture detection (at wrist and elbow joint) and assessment of ligament tear in rotator cuffs. With hospital resources being stretched during the COVID-19 pandemic, ultrasound is also seen as a valuable triage tool for detecting lung involvement in patients with COVID-19. In all these cases, the reliability of ultrasound examinations partly depends on sonographers' expertise in acquiring high quality images containing all necessary anatomical landmarks. Acquiring such 2D ultrasound (2DUS) images requires many hours of training with experts which is impractical in emergency care. Compared to 2DUS, 3D ultrasound (3DUS) with ultrasound videos cover a larger anatomical area and are generally easier to acquire for novice users\cite{ref_9}. However manual interpretation of these videos is tedious and time consuming. As a result, ultrasound remains a ubiquitous yet underutilized modality in emergency departments.

Automatic interpretation of ultrasound has been more challenging than Magnetic Resonance Imaging (MRI) and Computed Tomography (CT) due to noise artifacts, blurred anatomical boundaries and more importantly operator and system dependence. Specular reflective structures like wrist, elbow, hip, and pleura are even harder to automatically interpret due to the effect of beamwidth \cite{ref_10} and shadowing which significantly alters the pixel-intensities around the bone. Hence pixel intensity based approaches like intensity thresholding perform poorly in ultrasound images. Local phase (LP) information has been successfully used for detecting  robust features from B-mode ultrasound. But, these approaches generally require separate techniques for localization as there could be similar echogenic regions in close proximity, LP filtering would also lose some of the useful information in the original B-mode image.

Deep Learning (DL) models like convolutional neural networks (CNN) and  recurrent neural networks (RNN) have been used used for segmentation and classification in ultrasound images including Lung\cite{ref_11_A,ref_11_B} and Wrist\cite{ref_11_C}. Recently, CNN models that fuse LP filtered images with the original B-mode image have been used for bone segmentation \cite{ref_12}. All these DL models were trained using supervised learning techniques which depend on large fully annotated datasets. However, manual annotation  of ultrasound is tedious, subjective and time consuming in practice which results in scarcity of training data. Another limitation is the variability in ground truth annotation which  results in diminished performance of DL models. Most of the models are trained on data acquired from a single center and a single ultrasound scanner. Due to this lack of diversity in the training set, DL models might not generalize across different ultrasound scanners acquired by sonographers of varying levels of expertise. These fundamental limitations of supervised learning and the availability of large numbers of unlabeled data  has piqued research interest in unsupervised approaches in medical image analysis \cite{ref_12_A}.

These models exploit spatial \cite{ref_13} and temporal \cite{ref_14} dependencies in the data to generate low dimensional representations. Earlier approaches for unsupervised learning in medical images were based on clustering \cite{ref_14_A}. Moriya et al extended a joint unsupervised learning (JULE) framework to 3D medical images using a separate layer of 3D convolutions\cite{ref_14_A}. Semantic segmentation in natural images and videos has been approached using unsupervised techniques\cite{ref_14_B}. Li et al used instance segmentation embeddings learned from static images with  optical flow to segment moving objects in video\cite{ref_14_B}. Similarly, unsupervised key point detection techniques have also been proposed for natural images\cite{ref_16,ref_16_A,ref_16_B}. Unlike the unsupervised segmentation approaches, keypoint detection techniques offer an elegant framework to learn concise geometric representations \cite{ref_15}. Recently, Kulkarni et al proposed a transporter framework\cite{ref_15} for encoding fine-grained locations and orientations of object parts with potential advantages in reinforcement learning applications, this model automatically identifies key points in a video sequence by transporting features from a source frame at source frame's keypoint locations to target frame's keypoint locations for reconstruction of the target frame with the source frame. 

Inspired by the success of transporter networks for natural videos, an unsupervised key point detection framework for concise representation of locations and orientation of structures in ultrasound videos is proposed in this work. We develop a physics driven domain aware transporter framework and introduce local phase information as a feature probability map (FPM). Exploiting the limited range of orientations of significant features seen in ultrasound images, we also use the Radon Transform to highlight features within a customized range of  angular orientations. The  main contributions of this paper are :
\begin{enumerate}
\item A new attention mechanism in the transporter module using a  learnable keypoint heatmap in which each key point is weighted by a relevance score.
\item Integration of an FPM that allows the model to focus on specific regions in the image filtered by local phase at multiple scales. 
\item An acoustic feature fusion convolutional neural network (FF-CNN) to generate features relevant to ultrasound based on the FPM.
\item Use of radon transformed feature probability map (RT-FPM) at multiple orientations to highlight clinically relevant features 
\end{enumerate}
The rest of the paper is organized as follows. Section II presents the details of the proposed approaches. An exhaustive analysis and the respective results are presented in Section III followed by discussions and future work in Section IV and Conclusion in Section V. Finally, Conclusions are presented in Section VI.

\begin{figure*}[!htb]
\centering
\includegraphics[width=\linewidth]{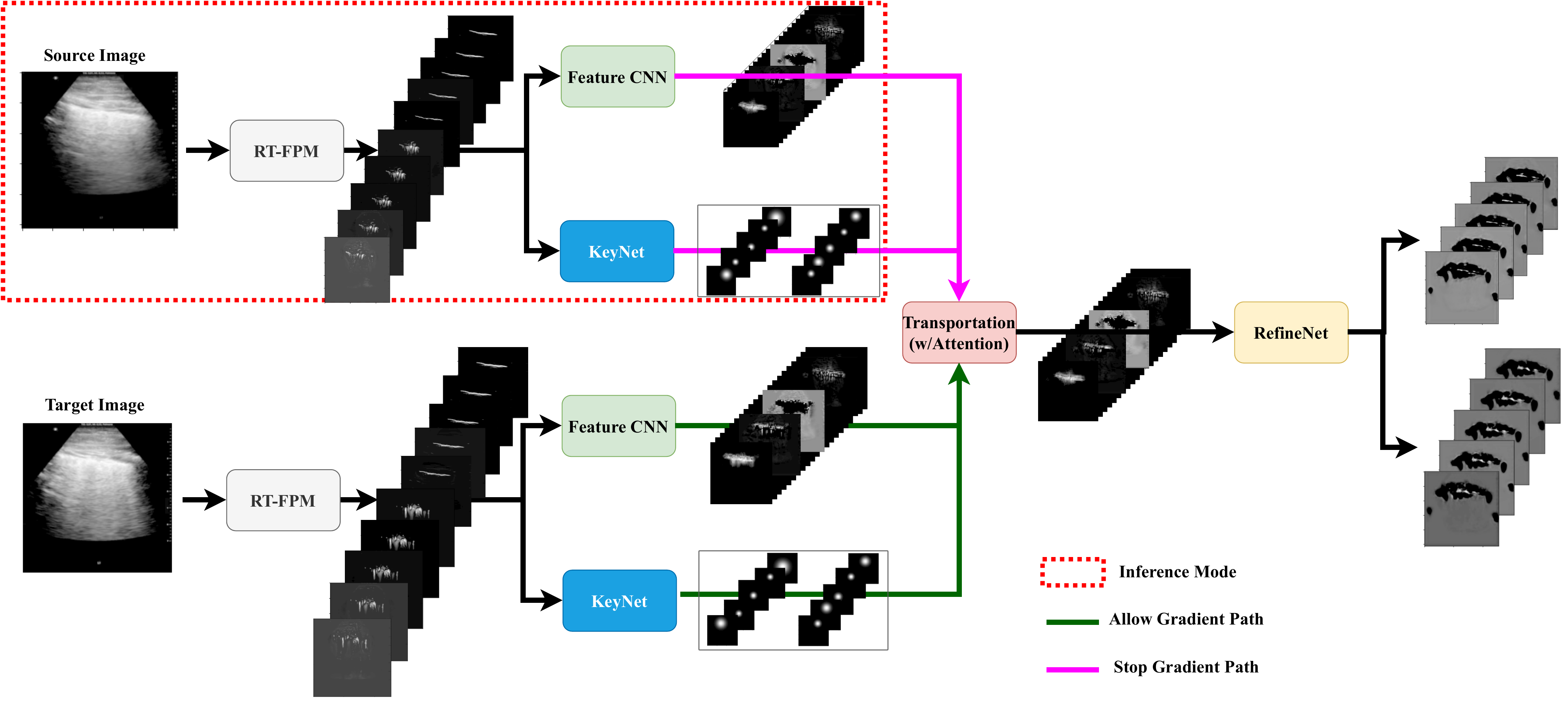}
\caption{Proposed framework consisting of DGA, radon transformed feature probability map (RT-FPM), FF-CNN \& KeyNet for generation of feature map \& point maps for both source and target images/frames, transporter architecture for transporting features at the key points from the target feature map to the source and a RefineNet to reconstruct the target frames} \label{fig1}
\end{figure*}

\section{Methods}
We identify $k$ keypoints for a given ultrasound frame pair ($x_{t}$,$x_{t+i}$) in a video $I_{t}$ during the training and for a single ultrasound frame during inference. As illustrated in Fig. \ref{fig1}, the key components are acoustic Radon Transformed Feature Probability Map (RT-FPM) of the frame pair, i.e, ($T\left(x_{t}, \lambda_{0}\right)$, $T\left(x_{t+i}, \lambda_{0}\right)$), a feature extraction CNN (FF-CNN) which provides embeddings corresponding to the FPM, a KeyNet which provides the keypoints which are transported to target images using a novel attention mechanism and finally target reconstruction using the RefineNet. Instead of a generic feature map extracted by a CNN directly from raw images (as used in \cite{ref_15}), we generate a domain aware feature map ($\Psi\left(x_{t}\right)$) using the feature probability of raw images which is processed by the Acoustic FF-CNN. The RT-FPM uses  distance-gain attenuation (DGA), local phase extraction at multiple scales and Radon transform to leverage the orientation of relevant structures. 

Formally, during the training phase, the model processes the radon transformed feature probability map (RT-FPM) of the frame pair ($T\left(x_{t}, \lambda_{0}\right)$, $T\left(x_{t+i}, \lambda_{0}\right)$) into FF-CNN extracted features ($\Psi\left(x_{t}\right)$, $\Psi\left(x_{t+i}\right)$) and KeyNet generated gaussian heat maps ($\Phi\left(x_{t}\right)$, $\Phi\left(x_{t+i}\right)$), however the gradient is allowed to backpropagate only through $\Phi\left(x_{t+i}\right)$ and $\Psi\left(x_{t+i}\right)$.  Similar to \cite{ref_15}, the model learns to predict the subsequent frame ($x_{t+i}$) represented by  $T\left(x_{t+i}, \lambda_{0}\right)$ based on the current frame ($x_{t}$) represented by $T\left(x_{t}, \lambda_{0}\right)$ in the sequence during evaluation. 

\subsection{Radon Transformed Feature Probability Map}
\begin{figure}[!htb]
\centering
\includegraphics[width=\linewidth]{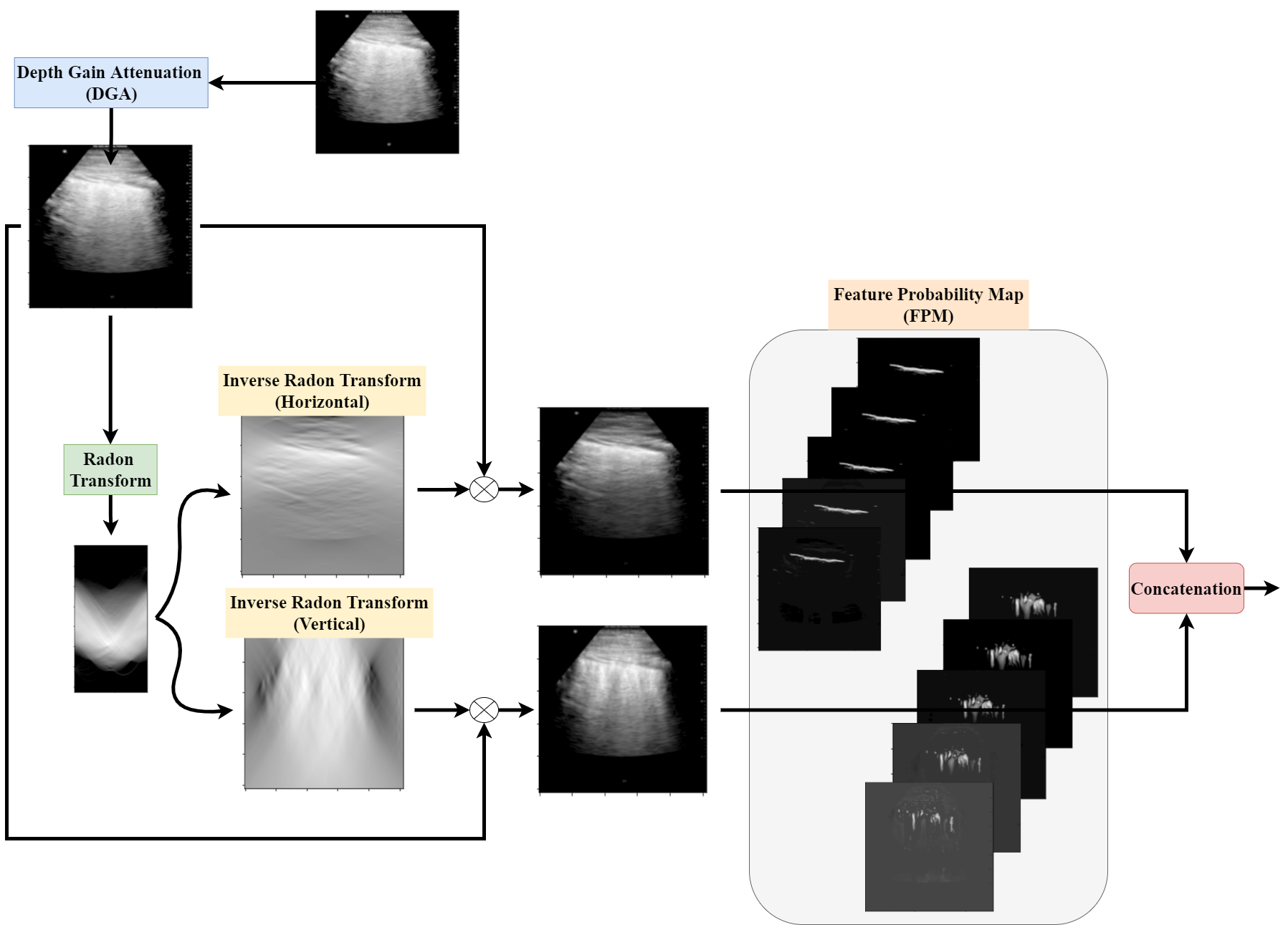}
\caption{Proposed RT-FPM framework consisting of DGA for artifact attenuation, RT-Controlled IRT approach for selective angle reconstruction and acoustic feature map generation} \label{fig_RT-FPM}
\end{figure}

The first step in the proposed approach is to do a DGA of the image, $x_{t}$, to suppress the spurious bright regions occurring due to muscular and fat regions in Lung and Wrist Ultrasound. The DGA applied to image $x$ can be represented as in $DGA(x)=\chi x$ where $\chi$ represents a depth dependent decay mask whose value at depth $d$, i.e, $\chi(d)=1-e^{-a \times d} / \max \left(e^{-a \times d}\right)$ depends on an exponential attenuation factor $a$. We then apply Radon Transformation (RT) followed by controlled inverse Radon Transform (IRT) as shown in Equations (\ref{eq1})-(\ref{eq2}) to extract angular features (within the interval $\theta_i$ to $\theta_f$) relevant to the given ultrasound domain as represented by $IR(i,j)$ in Equation (\ref{eq2}).

\begin{equation}
\label{eq1}
R(\rho, \theta)=\int_{-\infty}^{\infty} \int_{-\infty}^{\infty} f(i, j) \delta(i \cos \theta+j \sin \theta - \rho) d i d j
\end{equation}

\begin{equation}
\label{eq2}
IR(i, j)=\left.\int_{\theta_i}^{\theta_f} R(\rho, \theta)\right|_{\rho=i \cos \theta + j \sin \theta} d \theta
\end{equation}

where $f(i,j)$ represents the image intensity at pixels $(i,j)$, $\rho$ and $\theta$ represent the position and angle axes for the RT. In the proposed approach $\theta_i$ and $\theta_f$ are chosen to be $[0\degree -30\degree \& 150\degree-180\degree]$ for vertical edges and $[65\degree-115\degree]$ for horizontal edges respectively. The hypothesis behind the choice of these angular ranges is that the horizontal (pleura and A-lines in the case of LUS and bone in the case of WUS) and vertical (B-lines and B-patch in the case of LUS) features constitute the clinically relevant landmarks. The horizontal and vertical IRT images are further masked upon DGA compensated images to obtain enhanced horizontal and vertical features of the original ultrasound images as shown in Fig. \ref{fig_RT-FPM}.\\
With the enhanced horizontal and vertical components of the ultrasound image, we extract local phase information using a Gabor filter bank $G\left(x, \lambda_{0}\right)$ as shown in Equation (\ref{eq3}) where $\lambda_{0}$ is a scaling parameter that can be varied  as to capture image features at different resolutions, i.e, $\lambda_{0}$ varies between $0$, $4.2\pi\gamma_{k}$,  $8.4\pi\gamma_{k}$, $12.6\pi\gamma_{k}$, $16.8\pi\gamma_{k}$, where $\gamma_{k}$ = $3k$ for the $kth$ channel generated by both the enhanced vertical and horizontal radon components ($k=1$ to $5$) (refer Fig. \ref{fig_RT-FPM}). The value for $\sigma_{0}$ was fixed at 0.55.

\begin{equation}
\label{eq3}
G(\omega)=\exp \left(\frac{-\left(\log \left(\frac{\mid\omega \mid}{\omega_{0}}\right)\right)^{2}}{2\left(\log \left(\sigma_{0}\right)\right)^{2}}\right) \textrm{, where } \omega_{0}=\frac{2 \pi}{\lambda_{0}}
\end{equation}

Using monogenic signal analysis \cite{ref_11}, we model the image $I(x, y): I \in R^{2}$ as a combination of amplitude $A(x,y)$  and phase  as 

\begin{equation}
\label{eq4}
I(x, y)=A(x, y) \times \cos (\theta)
\end{equation}

The corresponding local phase filtered image can be written in tensor representation using symmetric features ($T_{even}$), asymmetric features ($T_{odd}$) and instantaneous phase $\varphi$ as:

\begin{equation}
\label{eq5}
L P T=\sqrt{T_{even}^{2}+T_{odd}^{2}} \times \cos (\varphi)    
\end{equation}

Tensors of symmetric and asymmetric features ($T_{even}$ and $T_{odd}$) are computed using Hessian $H$, Gradient $\nabla$, and Laplacian $\nabla^{2}$ operations as shown below:

\begin{equation}
\label{eq6}
\begin{array}{l}
T_{even}=\left[H\left(I_{B P}(x, y)\right)\right]\left[H\left(I_{B P}(x, y)\right)\right]^{T} \\
T_{odd}=-0.5 \times\left[\nabla\left(I_{B P}(x, y)\right)\right]\left[\nabla \nabla^{2}\left(I_{B P}(x, y)\right)\right]^{T}+\\\left[\nabla \nabla^{2}\left(I_{B P}(x, y)\right)\right]\left[\nabla\left(I_{B P}(x, y)\right)\right]^{T}
\end{array}   
\end{equation}

Three monogenic signals $M_{1}$, $M_{2}$ and $M_{3}$ are then calculated by applying the Riesz transform on LPT as described in \cite{ref_17} . Using these monogenic signals we define the Local Phase($LP(x)$), feature symmetry ($FS(x)$)  and pixel wise Integrated Backscatter map ($IBS(i,j)$):

\begin{equation}
\label{eq7}
\begin{array}{l}
LP(x)=1-\arctan \left(\frac{\sqrt{M_{2}^{2}+M_{3}^{2}}}{M_{1}}\right) \\
FS(x)=\frac{\max \left(T_{even}-T_{odd}-\tau\right)}{M_{1}^{2}+M_{2}^{2}+M_{3}^{2}}
\end{array}    
\end{equation}

$IBS(i, j)=\sum_{k=1}^{i} I^{2}(i,j)$ represents the integrated back scatterer map along the row direction. Finally these are combined to generate a feature probability map $T\left(x, \lambda_{0}\right)$:

\begin{equation}
\label{eq8}
T\left(x, \lambda_{0}\right)=LP(x) \times FS(x) \times(1-IBS(x))
\end{equation}

A detailed illustration of (\ref{eq8}) is presented in \cite{mahesh_icip}.

\subsection{Acoustic FF-CNN}
Once the domain specific  feature maps are extracted by the proposed RT-FPM, the acoustic FF-CNN is employed to encode the multiple scale and multiple orientation channels into lower dimensional embeddings. Since the input is the multi-channel multi resolution images, the acoustic FF-CNN ($\psi$) could also be considered as a multi-resolution encoding network. It is trained in an end-to-end manner with the whole network as described in the training section (II.G). The architecture of the FF-CNN consists of $5$ convolutional layers (all the CNN kernels are $3$x$3$ except for the kernel in first layer which is $7$x$7$) followed by batch normalisation and ReLU Activations. 

\begin{figure}[!htb]
\centering
\includegraphics[width=\linewidth]{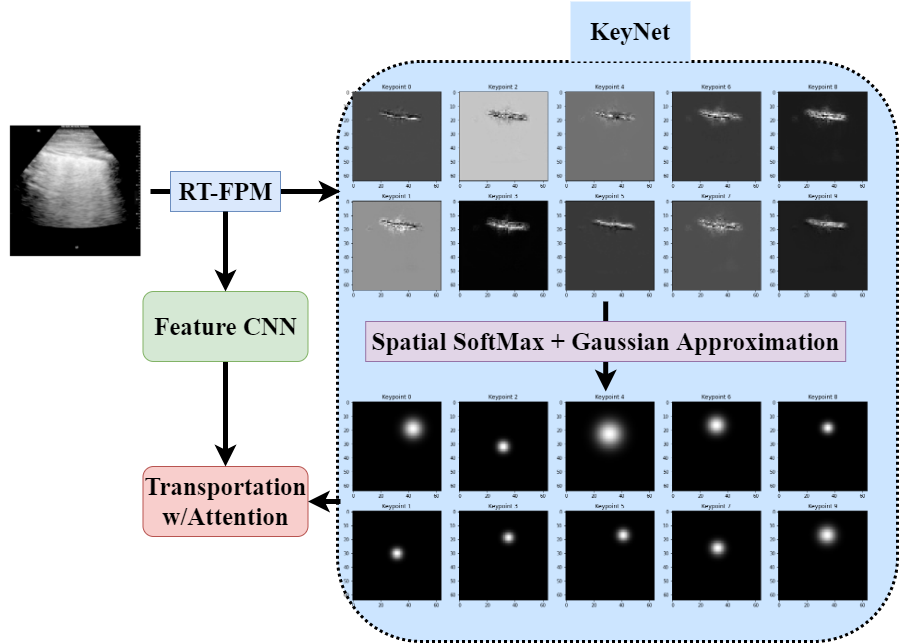}
\caption{The KeyNet network processes RT-FPM processed frames to extract landmark locations by performing convolution operations followed by Spatial Softmax and Gaussian approximation, the outputs of FFCNN and KeyNet for a frame pair are staged for training (refer Figure \ref{fig1}). The extracted keypoints in the figure are extracted by a trained Transporter with Attention network, note that a vanilla Transporter network would output equal sized keypoints instead.\\} \label{keynet}
\end{figure}

\subsection{KeyNet}
Similar to the Acoustic FF-CNN, KeyNet ($\phi$) also takes the RT-FPM as the input and acts as a multi-resolution encoding network. Further, Spatial Softmax is applied to the output followed by Gaussian heatmap estimation. These Gaussian heatmaps are interpreted as keypoints ($\phi(x_t)$) on the input image as shown in Fig. \ref{keynet}). It is trained in an end-to-end manner with the whole network as described in the training section (II.G). The architecture of the FF-CNN consists of $6$ convolutional layers (all the CNN kernels are $3$x$3$ except for the kernel in first layer which is $7$x$7$ and last layer which $1$x$1$) followed by batch normalisation and ReLU Activations. 

\begin{algorithm}	
	\caption{Transportation}
	\KwIn{\\
		KeyNet extracted source keypoints \bm{$\phi(x_{t})$}\;
		KeyNet extracted target keypoints \bm{$\phi(x_{t+i})$}\;
		FF-CNN extracted source features \bm{$\psi(x_{t})$}\;
		FF-CNN extracted target features \bm{$\psi(x_{t+i})$}\;	
		}			
	\KwOut{\\Transported feature map  \bm{$\epsilon(x_{t})$}\;}		
	\textbf{Initialization}:\\			
	\tcc{Transporting the features at each source keypoint location to corresponding target keypoint location}	
	$k \leftarrow len(\phi(x_{t})) \leftarrow len(\phi(x_{t+i}))$\;
	$\epsilon(x_{t}) \leftarrow \psi(x_{t})$\;

	\For{$k_{t},k_{t+i}$ in $\phi(x_{t}),\phi(x_{t+i})$}
	    {
	    $\epsilon(x_{t}) \leftarrow (1 - k_{t})*(1 - k_{t+i})*\epsilon(x_{t}) + k_{t+i}*\psi(x_{t+i}) )$\;
	    }
\label{alg:transportation}
\end{algorithm}

\begin{algorithm}	
	\caption{Loss function}
	\KwIn{\\
	    RefineNet reconstructed target frame \bm{$\eta(x_{t})$}\;
	    
		RT-FPM processed target frame \bm{$T\left(x_{t+i}, \lambda_{0}\right)$}\;
		}			
	\KwOut{\\Mean squared error for training the Transporter network \bm{$L(x_{t},x_{t+i})$}\;}		
	\textbf{Initialization}:\\		
	$l \leftarrow len(T\left(x_{t}, \lambda_{0}\right)) \leftarrow len(T\left(x_{t+i}, \lambda_{0}\right)) \leftarrow len(\eta(x_{t}))$\;
	$\epsilon(x_{t}) \leftarrow \psi(x_{t})$\;
	
	$L(x_{t},x_{t+i}) \leftarrow  \frac{\sum(\eta(x_{t}) - T\left(x_{t+i}, \lambda_{0}\right))^2}{dim(\eta(x_{t}))} $\;

\label{alg:lossfunction}
\end{algorithm}

\subsection{Transportation and Attention Mechanism}
As described in Algorithm 1, a transported feature map ($\epsilon\left(x_{t}\right)$) is created from the acoustic feature map of a source frame ($\Psi\left(x_{t}\right)$) and a target frame ($\Psi\left(x_{t+i}\right)$) by iterating over all $k$ keypoints in source frame and target frame, i.e, $\phi\left(x_{t}\right)$ and $\phi\left(x_{t+i}\right)$. This is done to suppress the features present at keypoint locations of source frame and target frame (described by the term $(1 - k_{t})\times{}{}(1 - k_{t+i})\times\epsilon(x_{t})$) and to transport features from the target frame (described by the term $k_{t+i}\times\psi(x_{t+i})$). In this work, two variants of attention mechanism in the transporter framework are introduced. This will allow the model to  adjust the strength of transportation capabilities of each of the $k$ keypoints:
\begin{itemize}
    \item Each keypoint possessing the the same standard deviation in their Gaussian approximation, but with adjustable degrees of transportation capability, i.e, keypoints having flexibility to partially transport features or enhance transported features. 
    \item Each keypoint possessing equal transportation capability of fully transporting features within their radius, but with adjustable standard deviation of Gaussian approximation. We initialize all keypoints with same standard deviation of $0.1$ with keypoints having the flexibility to adjust their standard deviation in order during the training process to transport major or minor geometric features. 
\end{itemize}

\subsection{RefineNet}
The RefineNet ($\eta$) is a generative model that outputs the reconstructed target frame ($\eta(x_{t}$) based on the transported feature map by inpainting regions in the neighbourhood of the keypoints. The RefineNet learns in an end-to-end manner with the minimization of the Mean Squared Error (MSE) loss function discussed in detail in the Section II.G. The input to RefineNet is the transported feature map $\epsilon(x_{t})$ generated by the outputs of FF-CNN and KeyNet as described in Algorithm 1. The architecture of the RefineNet consists of $3$ convolutional layers (all the CNN kernels are $3$x$3$) followed by batch normalisation and ReLU Activations. The upsampling is done using bi-linear interpolation operation. 

\subsection{Ultrasound Video Dataset}
In this multi center study, we retrospectively analyzed lung ultrasound data collected from Sree Chitra Tirunal Institute for Medical Sciences and Technology (SCTIMST), Trivandrum, India, Government Medical College Kottayam and Hospital Universitario Puerta de Hierro Spain and Wrist ultrasound data collected from University of Alberta Hospital, Edmonton, Canada.\\ 
\textbf{LUS Data:}
100 lung ultrasound videos taken from 40 subjects and using different ultrasound machines (GE Venue, Philips Lumify, Butterfly network and Fujifilm Sonosite). The ultrasound exam was performed with the patient in supine or near-supine position for the anterior scanning, and in the sitting or lateral decubitus position for the posterior scanning. The probe was positioned obliquely, along the intercostal spaces. The LUS examination was obtained moving the probe along anatomical reference lines, 2nd-4th intercostal space (ICS) of parasternal, midclavicular, anterior axillary and midaxillary line (on the right side to the 5th ICS), whereas for the posterior chest, the paravertebral (2nd- 10th ICS), sub-scapular (7th- 10th ICS) and posterior axillary (2nd- 10th ICS) lines. A video clip was recorded along each anatomical line, recording at least 3 seconds at each ICS. The examinations were performed using a phased and curvilinear array transducer (1.5–4.5 MHz). 
\textbf{WUS Data:}
113 3DUS videos (containing 256 frames each) were acquired from 57 children presenting at an emergency department with suspected unilateral fracture at the wrist. 3DUS images from both the affected and unaffected wrist along with conventional x-ray examination which was used to obtain gold standard diagnosis. Images were acquired on a Philips iU22 machine (using a 13-MHz 13VL5 probe for 3DUS) with the child seated in a neutral position. During examination, the injured wrist  was scanned on the dorsal (DS) and volar (VO) surfaces in both the sagittal and axial orientation resulting in four 3D scans of each wrist (DS sagittal,DS axial, VO sagittal and VO axial). While acquiring the 3DUS image the sonographer centered the view on the distal end of the radius in the different orientations. Each sweep was of 3.2 seconds duration through a range of +/-$15^{o}$ to with  ultrasound slices of 0.2mm. As a baseline, the same scanning protocol was followed for the unaffected wrist as well.  

\subsection{Training Details}
The Transporter architecture was implemented in PyTorch v1.7.1 and trained on 1024 pairs of images(each 256x256 pixels) for 100 epochs with Adam optimizer, learning rate = 0.001 decaying by 0.95/10 epochs and with a batch size as 16. The Gaussian heatmap standard deviation was initialised to 0.1. Other hyper parameters were adopted from \cite{ref_15} (refer to our source code at \url{https://github.com/tripathiarpan20/US-Transporter-eval}). 512 pairs of images were used for validation with 1 sample drawn per 10 frames. The network was trained to minimize the reconstruction loss, i.e, MSE between the RT-FPM extracted target frame reconstruction by the RefineNet and the RT-FPM extracted target frame (refer Fig. \ref{fig1}) as defined in Algorithm 2. As an ablation study, we also used a perceptual loss function between the target frame and reconstructed target frame using a pretrained VQ-VAE\cite{ref_13}. 

\subsection{Network Selection}
We used various encoder models for FF-CNN and KeyNet with identical DGA compensation and monogenic analysis. Highest accuracy in detecting fractures was obtained using  $6$ convolutional blocks for FF-CNN, $6$ convolutional blocks with a regressor for KeyNet and $6$ convolutional blocks for RefineNet with $7x7$ and $3x3$ convolutions with strides and padding varying between $1$ and $2$(refer to the supplementary material for details).
\section{Results}
The proposed approach was validated on both WUS and LUS datasets against ground truth. For the LUS dataset three radiologists identified features like A-lines, B-lines and pleura which are hallmarks of lung involvement and also classified the image frames into five classes on increasing severity of lung infection as in \cite{mahesh_icip}. Each subject was also categorised as normal or abnormal based on recorded clinical findings. 
Similarly for the WUS data, an expert sonographer identified the most common locations of fracture such as radial metaphysis, radial epiphysis and carpal bones. 
In the following subsections, the performance of the proposed RT-FPM and FF-CNN framework, keypoint detection with attention framework and the ability of the proposed approach for unsupervised detection and classification in LUS and WUS applications are discussed in detail.

\subsection{Performance of the proposed RT-FPM framework}
We progressively analyzed each aspect of the RT-FPM - DGA, acoustic feature map,Radon Transform and the convolutional block attention module (CBAM)

\subsubsection{Effect of DGA and acoustic feature maps}
\begin{figure}[!tb]
\centering
\includegraphics[width=0.65\columnwidth] {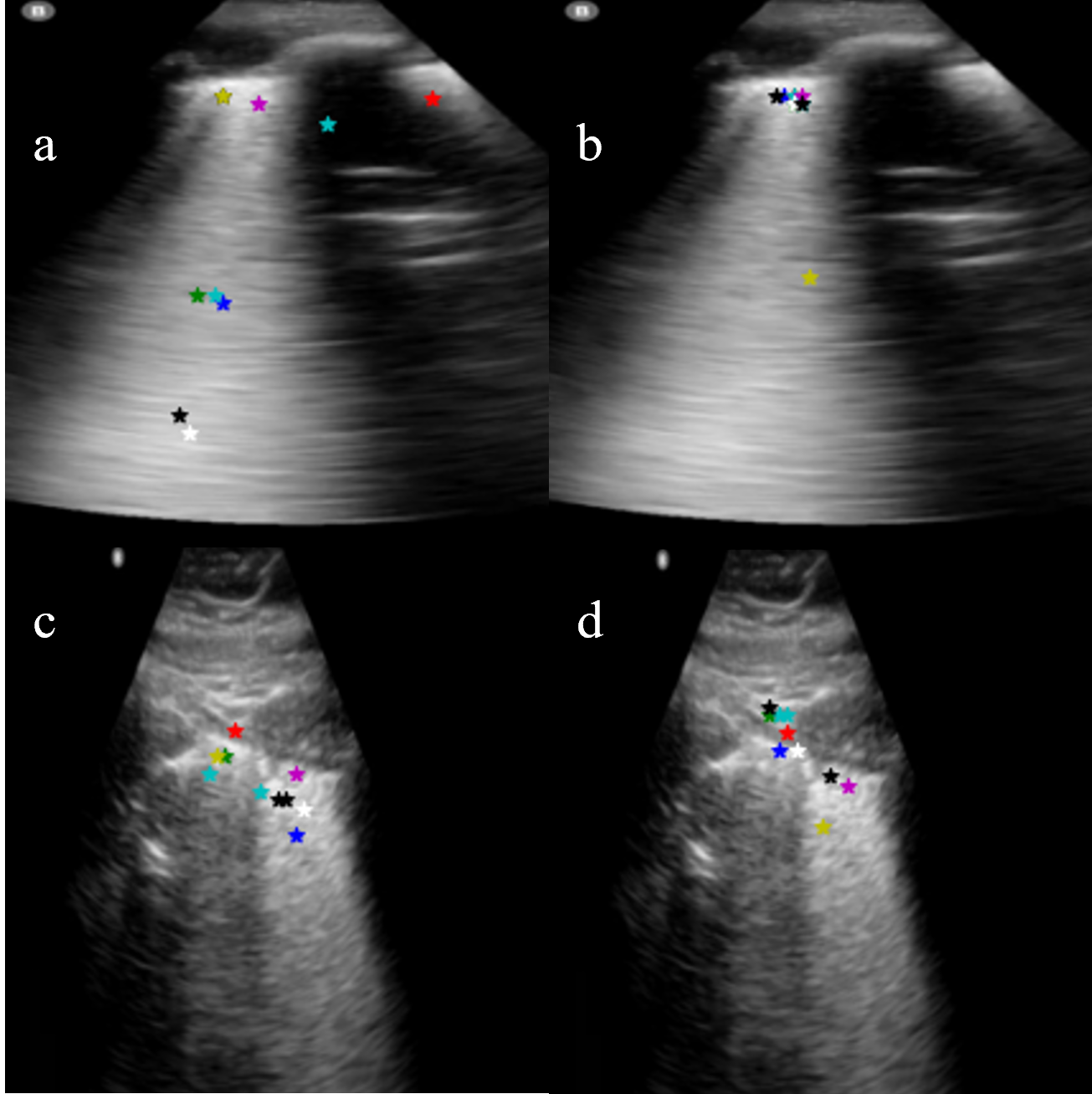}
\caption{(a, c) Without DGA, (b, d) With DGA. The results are far more stable in terms of pleural tracking in cases with bright B patches with the introduction of DGA.} 
\label{fig_DGA_ssim}
\end{figure}

As shown in Fig. \ref{fig_DGA_ssim} application of DGA gave a more stable distribution of key points. With DGA more points were identified on echogenic structures like the pleura.

\begin{figure}[!tb]
\centering
\includegraphics[width=0.65\columnwidth]{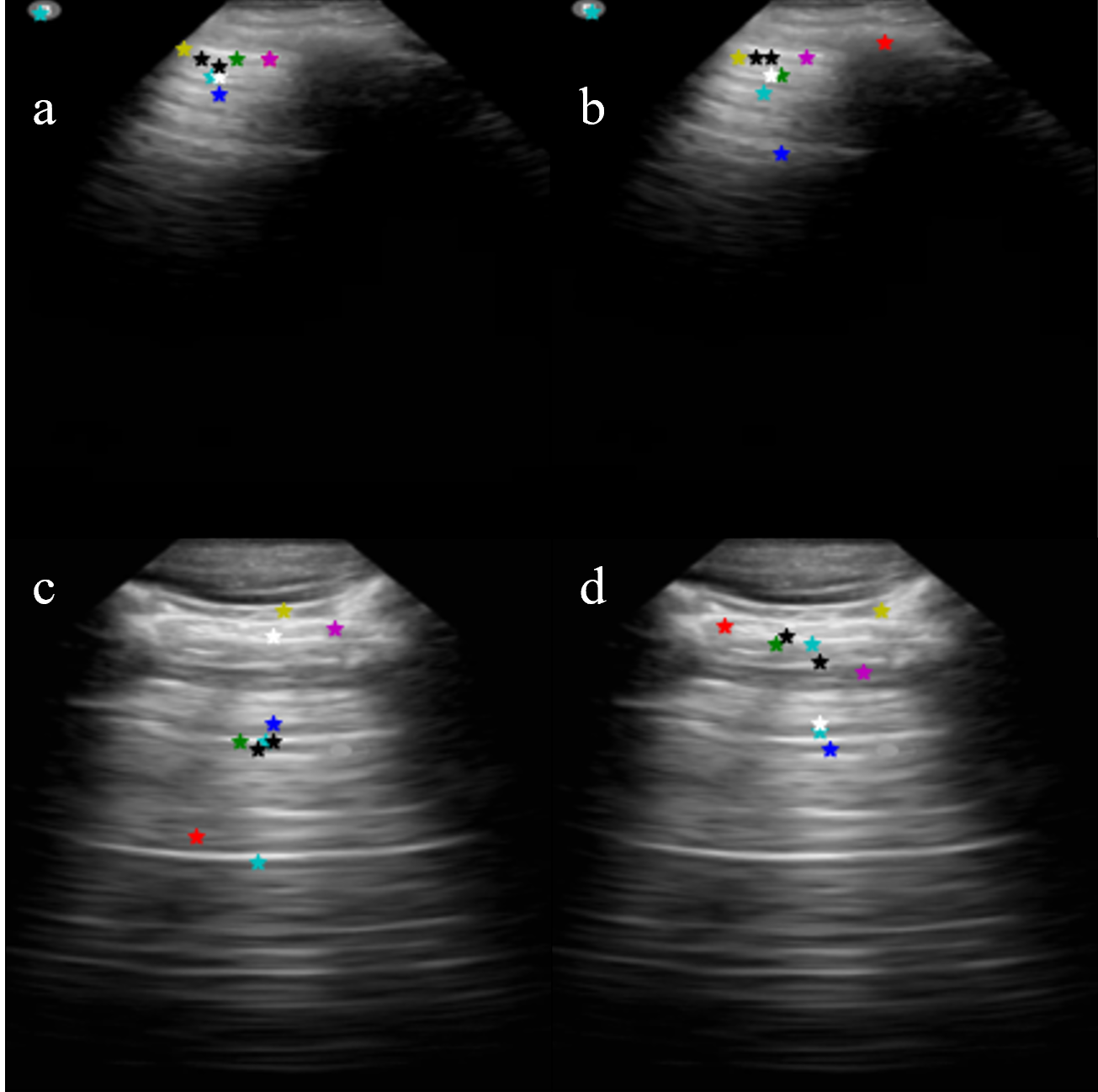}
\caption{(a, c) With normalized images by linearly varying ($\mu$, $\sigma$) parameters to form 10 input channels, (b, d) With acoustic feature fusion images by 10 varying wavelengths. The acoustic feature fusion (DGA and RT-FPM) is more closer to the desired landmarks (pleura in this case).} 
\label{fig_mu_sig}
\end{figure}

We compare the evaluation results obtained by preprocessing raw LUS frames into $10$ channels of varying wavelength acoustic feature fusion maps (formed by DGA and RT-FPM) and $10$ channels of grayscale images with varying levels of normalization, i.e, normalising the raw LUS frames by assuming a standard deviation ($\sigma$) and a mean ($\mu$) and clipping resultant pixel values to the range 0 and 1. For the latter, normalization with standard deviation ($\sigma$) of 0.5 and mean ($\mu$) varied linearly between $0.3$ and $0.7$ with respect to the index of the channel are employed to form $10$ input channels for the FF-CNN. The results for this ablation study, as shown in Fig. \ref{fig_mu_sig}, shows that the normalization process filters out irrelevant features of the LUS images and results are comparable with that of acoustic feature fusion. Acoustic feature fusion (DGA and RT-FPM) is  closer to the desired landmarks (pleura in this case). 

\subsubsection{Effect of Radon transform}
\begin{figure}[!tb]
\centering
\includegraphics[width=0.65\columnwidth]{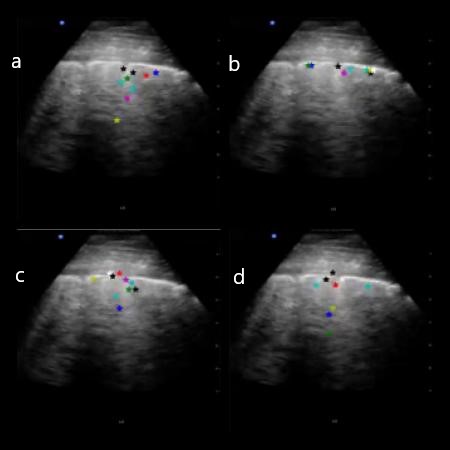}
\caption{(a) Without RT (b) With only horizontal IRT (c) With only vertical IRT (d) With both horizontal and vertical  RT} 
\label{fig_radon_ablation}
\end{figure}

We used the Radon transform to enhance  vertical and horizontal features and evaluated its effect on  keypoints. As seen in Fig. \ref{fig_radon_ablation}  the Radon transform guides the automatically detected key-points towards commonly seen echogenic features in the image without any specific prior information on the intensity or location of the feature. 

\begin{figure} [!tb]
\centering
\includegraphics[width=0.65\columnwidth] {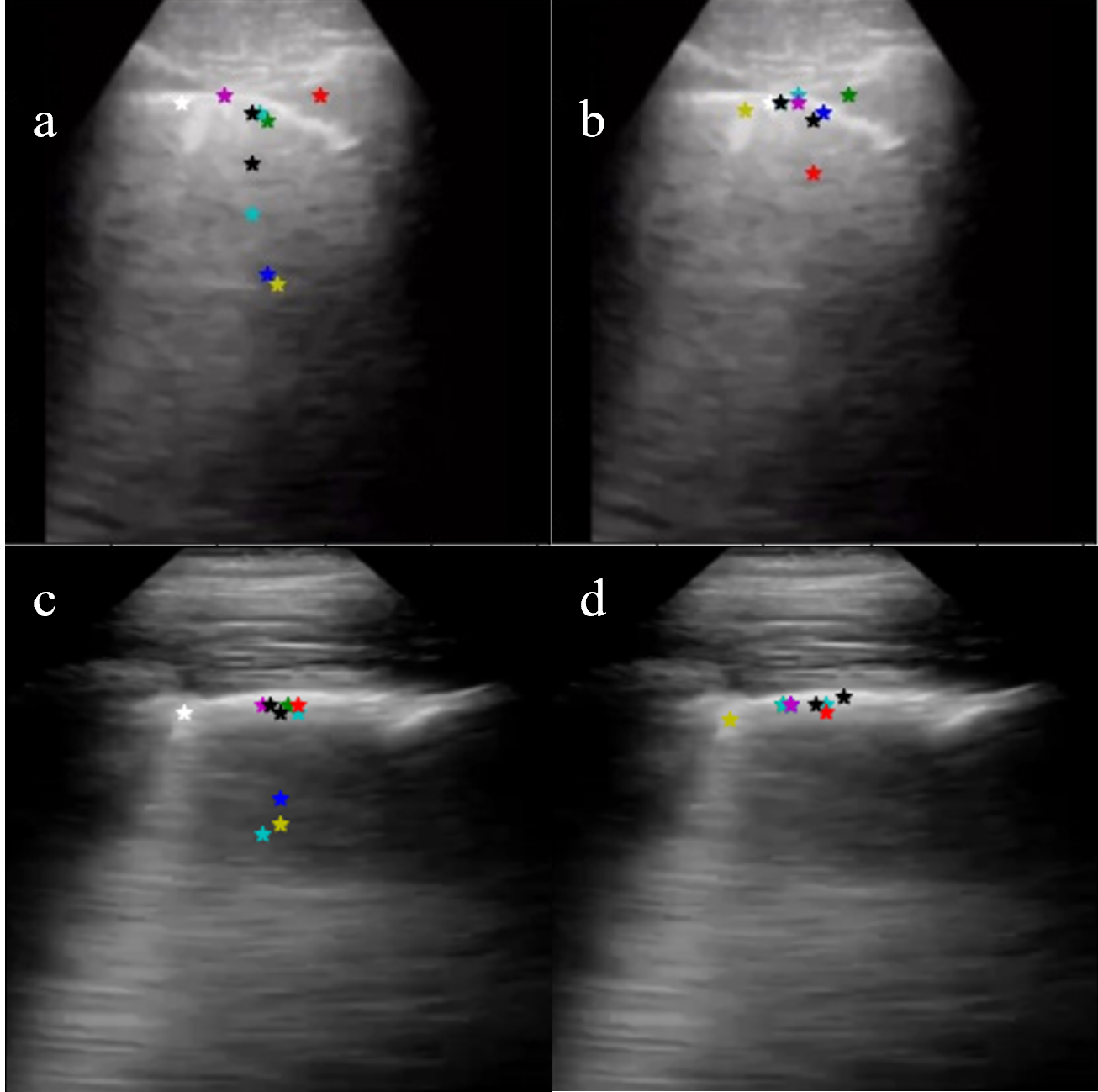}
\caption{(a, c) Without CBAM (b, d) With CBAM} 
\label{fig_cbm}
\end{figure}

\subsubsection{Effect of CBAM module}
We proposed a new convolutional block attention module  module \cite{CBAM} to filter irrelevant convolutional channels and spatial positions for better keypoint formation. As shown in Fig. \ref{fig_cbm}, application of  CBAM resulted in better tracking of  prominent features with keypoints without much cluttering. 

\subsection{Performance of the proposed keypoint detection with attention}
\begin{figure}
\includegraphics[width=0.48\textwidth]{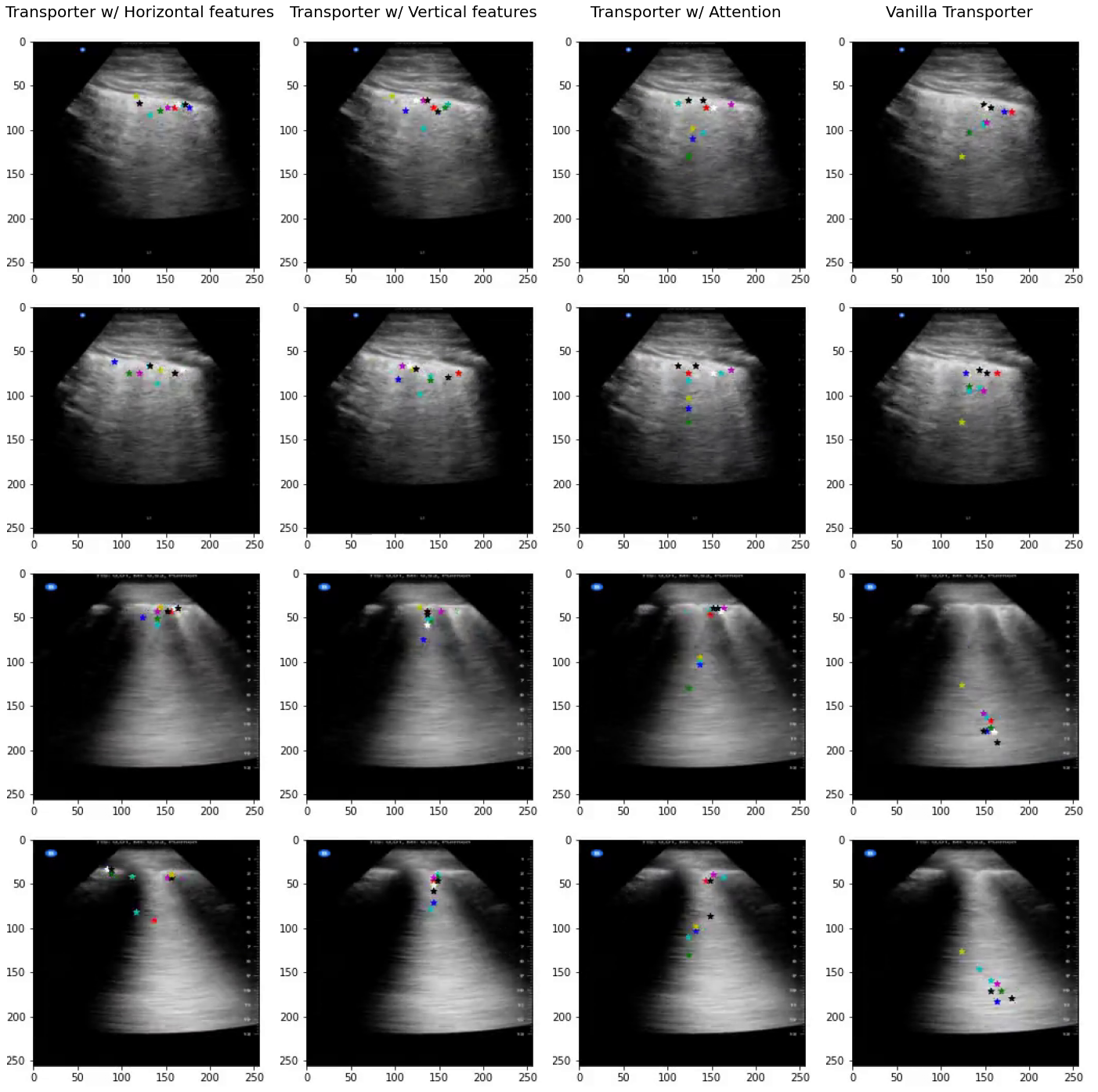}
\caption{Output of different Transporter networks on LUS frames} 
\label{fig_LUS_4x4}
\end{figure}

\begin{figure}
\includegraphics[width=0.48\textwidth]{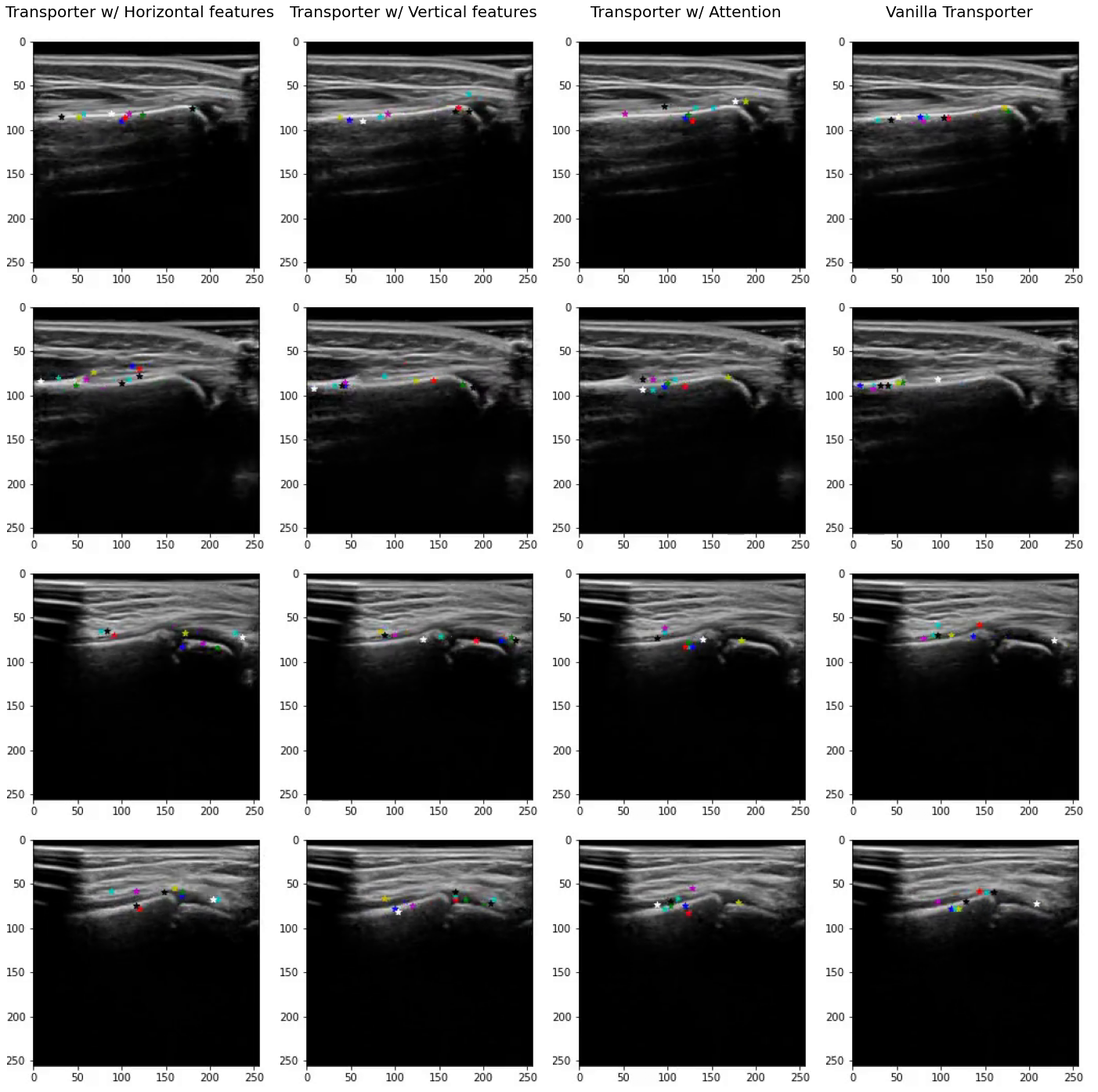}
\caption{Output of different Transporter networks on WUS frames} 
\label{fig_WUS_4x4}
\end{figure}

We evaluated the performance of the proposed keypoint detection with attention mechanism on the LUS and WUS datasets in terms of Specificity (SP) and Sensitivity (SN) in identifying relevant keypoints. The output of the transporter networks on LUS and WUS images are as shown in Figures \ref{fig_LUS_4x4} and \ref{fig_WUS_4x4}. The four columns in the plots represent the cases with only horizontal IRT, only vertical IRT, combined IRT with attention and the vanilla transporter outputs. It is evident from the figures that the proposed attention based scheme significantly improves the localization of keypoints within the desired landmarks. In WUS images, the  network also identified points near the bone discontinuity indicating fracture with out any explicit supervised training. 

In order to quantify the detection performance of the proposed approach, the estimated keypoints are compared against a manual segmented ground truth. In order to improve the performance of the keypoint detection particularly due to display labels, frame averaging has also been employed (note that frame averaging correction was not used during the training process). Frame averaging correction involves extracting average/median frame from the US video sequence and masking out the average/median frame from individual frames from the US video sequence, this correction helps in reducing the stationary but noisy artefacts from the US video frames. In fractured and non-fractured wrists, the proposed approach correctly identified the top portion of the bone with multiple keypoints (refer Figure \ref{fig_WUS_4x4}). In fractured wrists (third row in Figure \ref{fig_WUS_4x4}, the network was able to track keypoints near the fracture. The performance metrics on 58 manually labelled WUS videos are shown in Table \ref{tab:WUS_metrics}. In the case of LUS, binary segmentation data of LUS frames with landmarks like Pleura, A lines and B lines was used to compare the  performance metrics. The proposed approach detected these landmarks efficiently and the performance metrics on 1000 manually labelled LUS frames are shown in Table \ref{tab:LUS_metrics}. Additionally, to quantify the performance of the proposed unsupervised key point learning approach on Lung Ultrasound data further, the accuracy of detecting most common landmark, i.e, pleura, has been evaluated. The results show that the proposed approach has been successful in correctly detecting pleura in 950 frames out of 1081 frames which shows an accuracy of 91.8\%. The network was also evaluated on LUS segmentation data by considering the GradCam heatmap generated by KeyNet to be predicted segmentation with the Dice coefficient for pleural line detection, which was obtained to be 0.87.

\begin{table}[]
\centering
\caption{Evaluation metrics for Wrist Ultrasound on manually labelled segmentation data with Transporter having k=10, SP refers to the fraction of keypoints that detect a landmark and SN refers to the fraction of landmarks detected by at least one keypoint}
\label{tab:WUS_metrics}
\resizebox{\columnwidth}{!}{%
\begin{tabular}{|c|c|c|c|c|}
\hline
 & \begin{tabular}[c]{@{}c@{}}Vanilla \\ Transporter\end{tabular} & \begin{tabular}[c]{@{}c@{}}Transporter \\ w/ Attention\end{tabular} & \begin{tabular}[c]{@{}c@{}}Vanilla \\ Transporter\\ w/ frame avg\end{tabular} & \begin{tabular}[c]{@{}c@{}}Transporter \\ w/ Attention \\ w/ frame avg\end{tabular} \\ \hline
\begin{tabular}[c]{@{}c@{}}Specificity \\ (SP)\end{tabular} & 0.47 & 0.5 & 0.69 & 0.72 \\ \hline
\begin{tabular}[c]{@{}c@{}}Sensitivity \\ (SN)\end{tabular} & 0.69 & 0.7 & 0.66 & 0.74 \\ \hline
\end{tabular}%
}
\end{table}

\begin{table}[]
\centering
\caption{Evaluation metrics for Lung Ultrasound on manually labelled segmentation data with Transporter nets having k=10, SP refers to the fraction of keypoints that detect a landmark and SN refers to the fraction of landmarks detected by at least one keypoint}
\label{tab:LUS_metrics}
\resizebox{\columnwidth}{!}{%
\begin{tabular}{|c|c|c|c|c|}
\hline
 & \begin{tabular}[c]{@{}c@{}}Vanilla \\ Transporter\end{tabular} & \begin{tabular}[c]{@{}c@{}}Transporter \\ w/ Attention\end{tabular} & \begin{tabular}[c]{@{}c@{}}Vanilla \\Transporter\\ w/ frame avg\end{tabular} & \begin{tabular}[c]{@{}c@{}}Transporter \\ w/ Attention \\ w/ frame avg\end{tabular} \\ \hline
\begin{tabular}[c]{@{}c@{}}Specificity \\ (SP)\end{tabular} & 0.70 & 0.76 & 0.79 & 0.83 \\ \hline
\begin{tabular}[c]{@{}c@{}}Sensitivity \\ (SN)\end{tabular} & 0.96 & 0.99 & 0.97 & 0.99 \\ \hline
\end{tabular}%
}
\end{table}
From Tables \ref{tab:WUS_metrics} and \ref{tab:LUS_metrics}, it is clear that Transporter with attention has significant effect on the model performance. However, its significance is more apparent in the cases where the model is tasked with detection of a large number of keypoints. In such a scenario, the vanilla Transporter model overfits on the dataset leading to noisy keypoints. On the contrary, the Transporter with Attention network has the capability to suppress redundant keypoints while retaining the non redundant keypoints. This is achieved by adjustment of the standard deviation of the Gaussian approximation individually for all keypoints during the training process. 

\subsection{Unsupervised classification using the proposed keypoint detection with attention}
\begin{figure}
\includegraphics[width=\columnwidth]{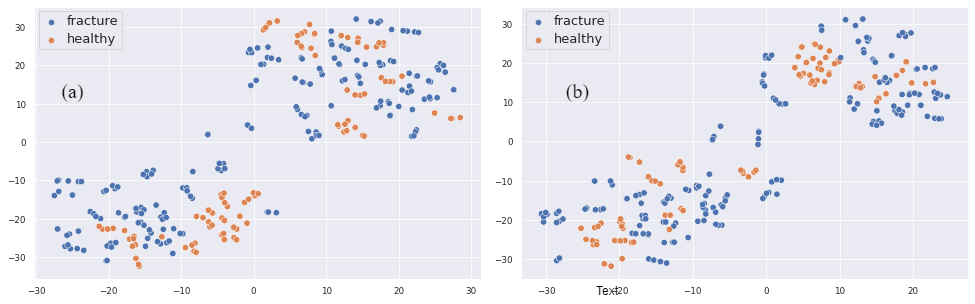}
\caption{t-SNE plots for Transporter networks on WUS video data. (a) without Attention (b) with Attention.} 
\label{fig_tSNE_WUS}
\end{figure}

\begin{figure}
\includegraphics[width=\columnwidth]{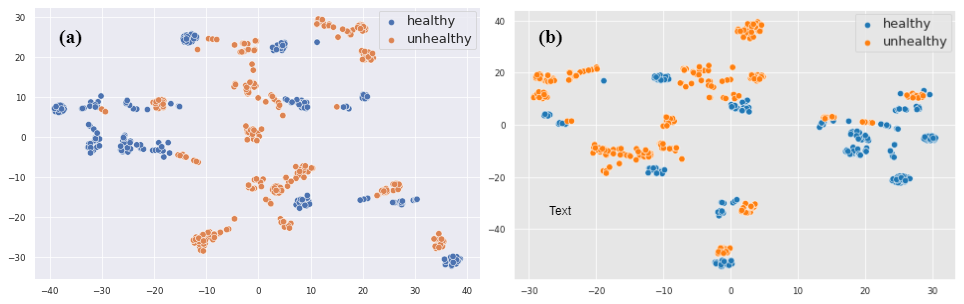}
\caption{t-SNE plots for Transporter networks on LUS video data. (a) without Attention (b) with Attention.} 
\label{fig_tSNE_LUS}
\end{figure}

\begin{figure}
\centering
\includegraphics[width=\columnwidth]{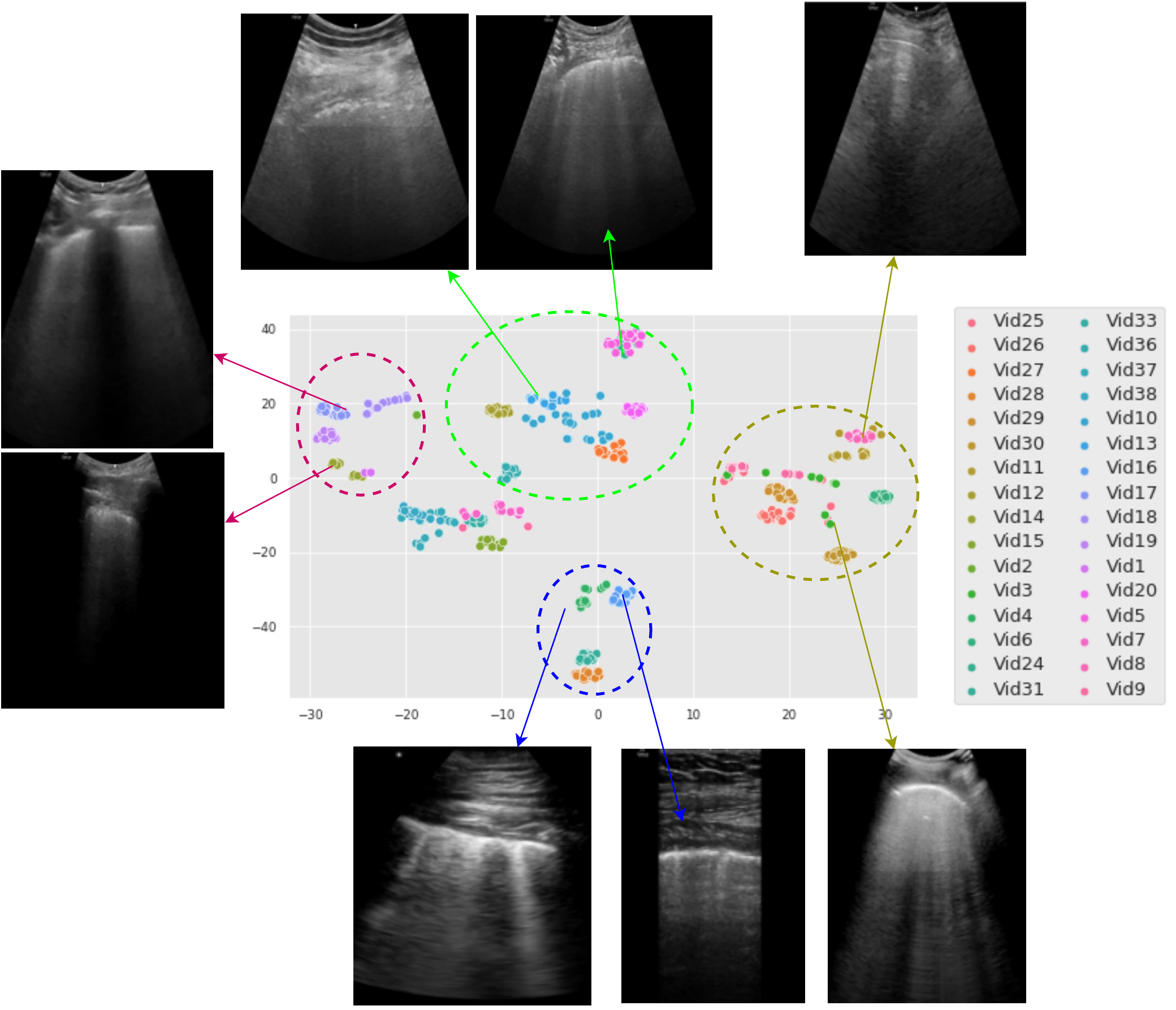}
\caption{ t-SNE plot on LUS data with  FF-CNN  embedding  of  Transporter  with  Attention  with frame sample to video instance labels. The plot clearly shows the clusters being formed based on frame similarity which shows the potential application of the proposed approach for unsupervised clustering of video frames based on similarity.}
\label{fig_tSNE_LUS_Vid}
\end{figure}

\begin{figure}
\centering
\includegraphics[width=\columnwidth]{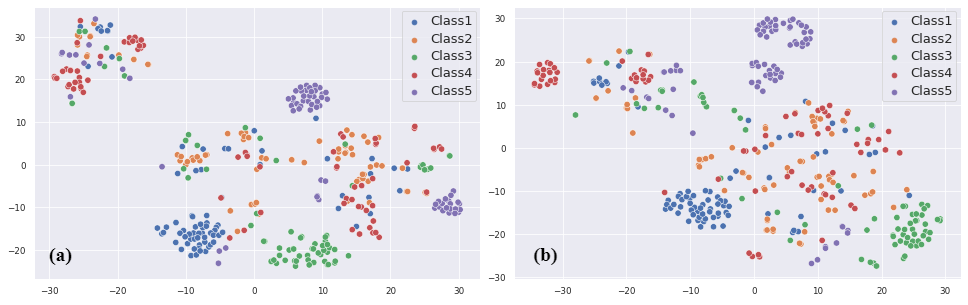}
\caption{tSNE plots with FF-CNN embeddings on LUS data with labels based on features, namely: Class 1: presence of only A lines below pleura, Class 2: presence of A and B lines together, Class 3: Only few scattered B lines (No A lines), Class 4: Patchy B lines (No A lines) and Class 5: Dislocated pleura without any visible A or B lines. (a) without Attention and (b) with Attention.} 
\label{fig_tSNE_LUS_5Class}
\end{figure}

In order to evaluate the possibility of employing the proposed approach for unsupervised classification applications, three separate classification scenarios are considered: $1)$ fracture vs non-fracture in the case of WUS, $2)$ normal vs abnormal lung for LUS and $3)$ various lung landmarks and manifestations divided into five classes as in \cite{mahesh_icip}. To explain the performance of the unsupervised classification, t-Distributed Stochastic Neighbor Embedding (t-SNE) plots \cite{tsne} are employed as below. In all the cases, the selected frames randomly drawn from $10$ experiments are forward passed through the Transporter in inference mode (refer Figure \ref{fig1}). The resultant FF-CNN embedding and labels were stored and used for plotting t-SNE for the median plot out of 10 experiments:
\begin{itemize}
    \item For fracture detection on WUS videos, $10$ experiments were conducted employing $58$ labelled videos (consisting of $30-50$ frames each) with $5$ frames randomly drawn from the interval of interest in each video. The corresponding t-SNE plots for with and without attention are as shown in Fig. \ref{fig_tSNE_WUS}. Although imperfect due to lack of labels during training, the plots show reasonable non-linear boundaries of separation between fracture and non-fracture cases. Also, the approach with attention appears to give clusters with lower standard deviation when compared to one without attention.
    \item For detection of healthy and unhealthy Lung US samples, 20 frames from 32 Lung US video sequences (consisting of 200-400 frames each) with 14 healthy and 18 unhealthy instances were drawn randomly in each of the 10 experiments. The corresponding t-SNE plots for with and without attention are as shown in Fig. \ref{fig_tSNE_LUS}. It is interesting to note that the t-SNE plot with attention employed shows appreciable separation of healthy and unhealthy classes even without any explicit training. Additionally, Figure \ref{fig_tSNE_LUS_Vid} demonstrates the inter and intra video frame correlation learned by the FF-CNN of the vanilla Transporter network.
    \item For the classification of LUS frames into five classes as in \cite{mahesh_icip}, 80 frames from each of the 5 classes from LUS dataset (each Class having 1000-1500 frames) were drawn randomly in each of the 10 experiments. The plots as shown in \ref{fig_tSNE_LUS_5Class} show appreciable non-linear boundary of separation for 3 out of 5 classes. The identified three classes are the one with only A-lines, only B-patch and pleural consolidations. The random distribution of points in the remaining 2 classes could be attributed to the co-existence of A-lines and B-lines. 
\end{itemize}

The t-SNE embeddings were also further treated as the representations of the LUS frames and were split into train and test set in the ratio of 70:30 and test split was classified with k Nearest Neighbour method. Such an approach can be used to classify new samples on the fly by applying the same non-linear transformation on the new points \cite{LION-tsne},\cite{online-tsne} as used in our experiments during the t-SNE as a co-classification method. The results for the same are reported in Table \ref{tab:LUS_co_class}. It is encouraging to note that without any explicit supervised training, the proposed physics driven domain specific transporter network with attention mechanism gives results comparable to that of a supervised classifier. 

\begin{table}[]
\centering
\caption{Comparison  of  classification  performance  of  the proposed unsupervised approach versus supervised approaches}
\label{tab:LUS_co_class}
\resizebox{0.95\columnwidth}{!}{%
\begin{tabular}{|c|c|c|c|c|c|}
\hline
\multicolumn{2}{|c|}{} & Accuracy & Precision & Recall & F1-score \\ \hline
\multirow{2}{*}{\begin{tabular}[c]{@{}c@{}}Normal\\ /\\ Abnormal Lung\end{tabular}} & Supervised & 0.91 & 0.85 & 0.98 & 0.91 \\ \cline{2-6} 
 & \begin{tabular}[c]{@{}c@{}}Proposed\\ Unsupervised \\ Co-classification\end{tabular} & 0.97 & 0.95 & 0.96 & 0.95 \\ \hline
\multirow{2}{*}{\begin{tabular}[c]{@{}c@{}}5 Class Lung \\ Scoring\end{tabular}} & Supervised & 0.86 & 0.82 & 0.91 & 0.86 \\ \cline{2-6} 
 & \begin{tabular}[c]{@{}c@{}}Proposed\\ Unsupervised \\ Co-classification\end{tabular} & 0.71 & 0.74 & 0.89 & 0.71 \\ \hline
\end{tabular}
}
\end{table}

\section{Conclusion and Future Scope}
We described a transporter neural network with components tailored to domain specific features seen in ultrasound. The proposed model has been validated on LUS and WUS datasets collected at multiple locations employing diverse ultrasound systems. Instead of using end-to-end neural networks, we introduced features specific to ultrasound to generate a feature probability map from the original ultrasound images. A DGA compensation was employed to suppress most of the  unwanted superficial reflections due to fat and muscular interfaces and to highlight the relevant structures located deeper in the image. Conventional DL models in ultrasound use supervised learning which requires precisely labeled ground truth annotations. The proposed keypoint detection framework uses unsupervised learning and relies on information from neighbouring frames. The transporter framework is robust to variation in the number and size of image landmarks commonly seen in ultrasound images. With a new attention mechanism we were able to reduce the number of redundant key points and improve its sensitivity in identifying imaging landmarks identified by the experts. Using the Radon Transformed Feature Probability Map (RT-FPM) we ensure that the neural network learns features and keypoints that are relevant to the applications. In this work, we have tuned the angular parameters to filter out relevant features in Lung and Wrist ultrasound images. With appropriate modifications the Radon transform approach can be adapted to other ultrasound examinations. \\
Visualization of the keypoint maps using t-SNE showed a clear differentiation between normal and fractured cases in WUS and normal and abnormal lung in LUS. In clinical practice automatic detection of key points provides a succinct summary of the ultrasound examination as it indicates the locations of abnormalities. Since the learning technique is task agnostic it can be potentially repurposed to other ultrasound use cases like elbow, hip and rotator cuff images that are interpreted based on certain well defined landmarks. For instance, it can be adapted to analyse ultrasound sequences of elbow or shoulder to detect fracture or ligament tears. 
Our study has limitations, as in our tracking technique we have not specifically addressed motion artifacts that are commonly seen in ultrasound scans. Although in most cases these artefacts occupy higher frequencies they could potentially affect the reconstruction of the feature map and result in incorrect keypoints. This can be addressed by associating spatial and temporal saliency features to each keypoint. \\
As a future work, we plan to incorporate a classification head in the transporter architecture which could be trained to detect fractures reported by human readers or landmarks in the lung to identify an abnormal frame. Towards this, it is envisaged to develop a hybrid loss function which combines MSE and cross entropy for frame reconstruction and binary classification. Features learned from the generative RefineNet model could be used to initialize the new classification head. This formulation also shows promise for obtaining more relevant domain specific keypoints for ultrasound data sets conditioned on the binary labels to be predicted by a classification head on top of FF-CNN features using gradient backpropagation through the keypoint locations. For instance, specialised keypoint detection for COVID-19/viral infections given their binary labelled lung ultrasound dataset with abnormal/normal classes , or specialised keypoint detection for fracture detection given binary labelled wrist/knee ultrasound dataset with fracture/healthy classes. In the above formulation, DGA might not be required due to the availability of labelled data which would impose sufficient control on the architecture during the training process to extract keypoints relevent to the labels. Additionally, DGA might not be required for a large value of k in the unsupervised Transporter, i.e, number of keypoints detected by KeyNet as the main purpose of DGA was to discard features that were auxillary so that existing keypoints focus more on important features, this claim remains unproven currently as we restricted our studies to a small number of keypoints (10-20) to focus more on remaining hyperparameters of the network. A trained Transporter architecture could also be used to formulate a reliable structural similarity index for a frame pair specific to ultrasounds (as inferred by Figure \ref{fig_tSNE_LUS_Vid}) by setting a reconstruction loss threshold in the Transporter network such that the reconstruction loss obtained would be below a threshold if the frame pair is sufficiently similar to each other and vice-versa.


\section*{Acknowledgment}
The authors would like to acknowledge funding from the Department of Science and Technology - Science and Engineering Research Board (DST SERB (CVD/2020/000221)). Authors would also like to acknowledge the computing resources at Compute Canada Cluster and NVIDIA and CDAC for PARAM SIDDHI AI System.

\section*{Supplementary Material}
The implementation of the network is made open-sourced and available publicly at \href{https://github.com/tripathiarpan20/US-Transporter-eval}{https://github.com/tripathiarpan20/US-Transporter-eval}.

\end{document}


\maketitle

\section*{Code and Data Availability}
In order to enable the reproducability of the proposed keypoint detection, the relevant inference scripts are made available freely at our official repository: \href{https://github.com/tripathiarpan20/US-Transporter-eval}{https://github.com/tripathiarpan20/US-Transporter-eval}.

\section*{Comments and Disclaimer:} 
\begin{enumerate}
    \item As mentioned in the main text, the present work demonstrates a prototype for a robust unsupervised ultrasound video key point pipeline. At present the system is trained and validated with ultrasound scans from various geographies obtained from different ultrasound machines by separate clinicians. Future work would include analyzing the proposed system with better pruned US scans i.e. data from distinct machine vendors, standardized ultrasound scans with specific presets which are expected to further increase the performance of the proposed methodology. 
    
    \item The following prototype is in development and should not be used as a substitute for a trained professional in making diagnostic decisions.
    
\end{enumerate} 

\section*{Supplementary Figures}
This section contains the additional figures which shows the architecture of FF-CNN, KeyNet, RefineNet, effect of the controlled Radon transform on LUS and WUS images and the expanded version of t-SNE plot for video-wise clustering.

\begin{figure*}[!h]
\centering
\includegraphics[width=\textwidth]{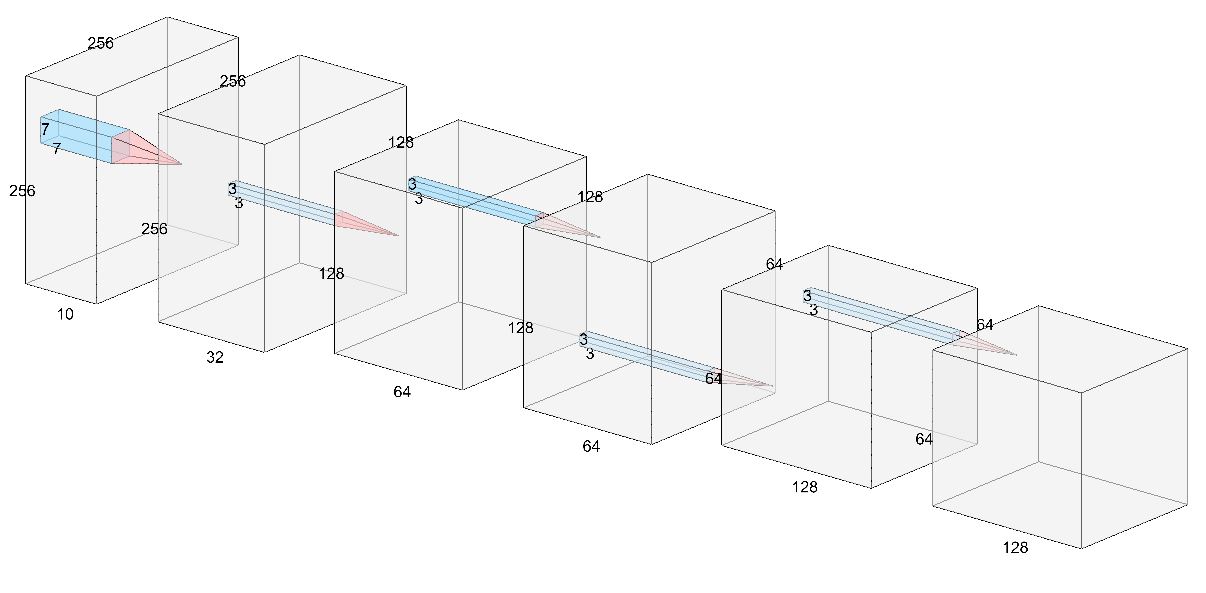}
\setlength{\belowcaptionskip}{-10pt}
\caption{Architecture for the FF-CNN in Transporter network, each block consists of Convolution operation followed by Batch Normalisation and ReLU Activation} 
\label{app_ffcnn}
\end{figure*}

\begin{figure*}[!h]
\centering
\includegraphics[width=\textwidth]{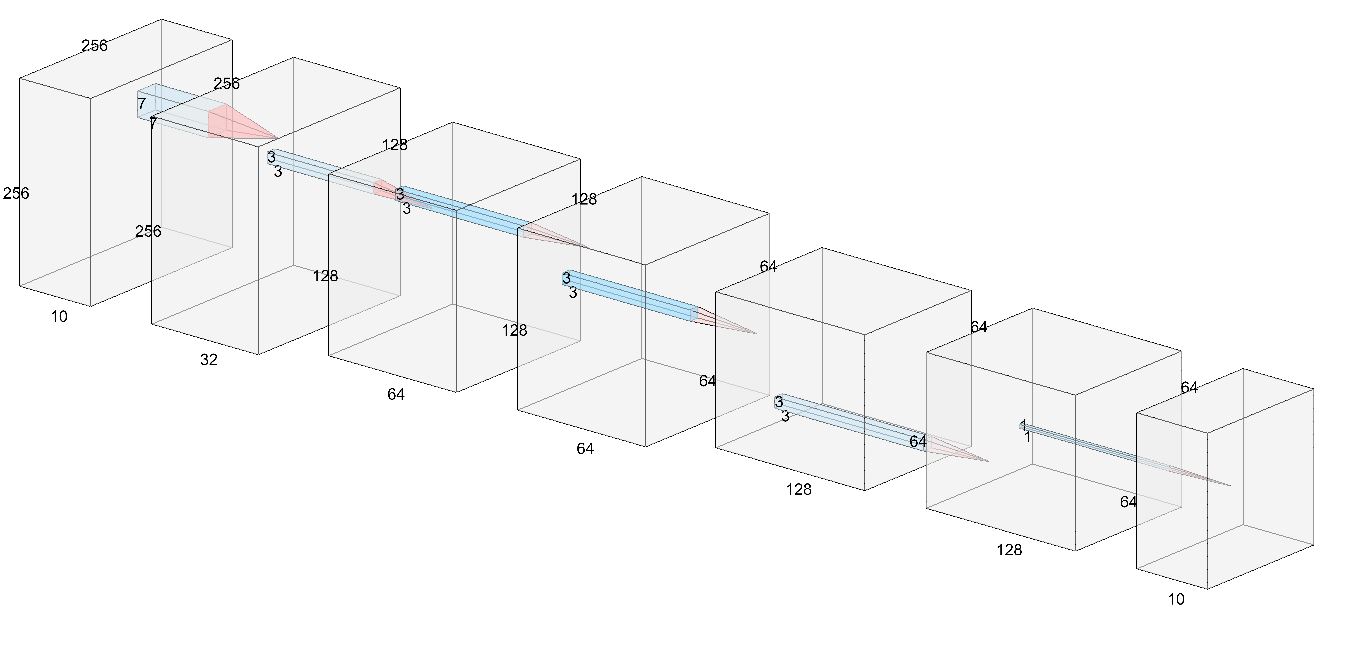}
\setlength{\belowcaptionskip}{-10pt}
\caption{Architecture for the KeyNet CNN for k=10 in Transporter network, each block consists of Convolution operation followed by Batch Normalisation and ReLU Activation, the final output of the CNN is further used for Gaussian approximation as seen in Figure 3} 
\label{app_}
\end{figure*}

\begin{figure*}[!h]
\centering
\includegraphics[width=\textwidth]{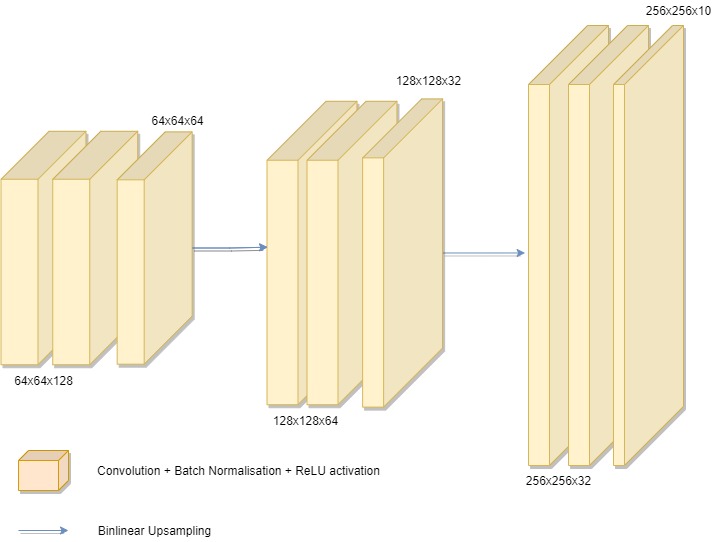}
\setlength{\belowcaptionskip}{-10pt}
\caption{Architecture for the RefineNet for k=10 in Transporter network, the input to the network is the transported feature map obtained with Algorithm 1, each block consists of Convolution operation followed by Batch Normalisation and ReLU Activation, the final output of the KeyNet is considered as the reconstruction of the representation of target frame, i.e, $T\left(x_{t+i}, \lambda_{0}\right)$}
\end{figure*}

\begin{figure*}[!h]
\centering
\includegraphics[width=\textwidth]{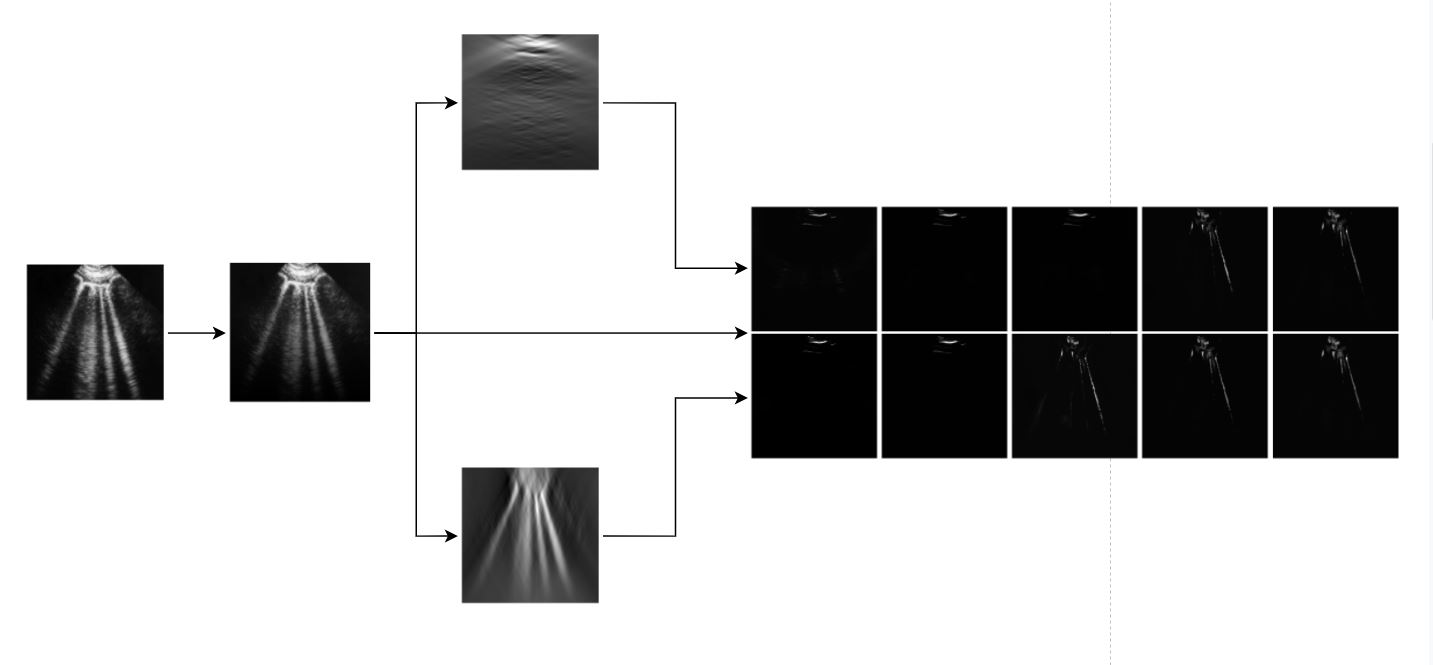}
\setlength{\belowcaptionskip}{-10pt}
\caption{DGA + RT-FPM preprocessing pipeline (refer Figure 1) of Lung Ultrasound frame with Septal Rockets, note that DGA attenuates the lower section of the raw frame in this case}
\label{fig_LUS_preprocess}
\end{figure*}

\begin{figure*}[!h]
\centering
\includegraphics[width=\textwidth]{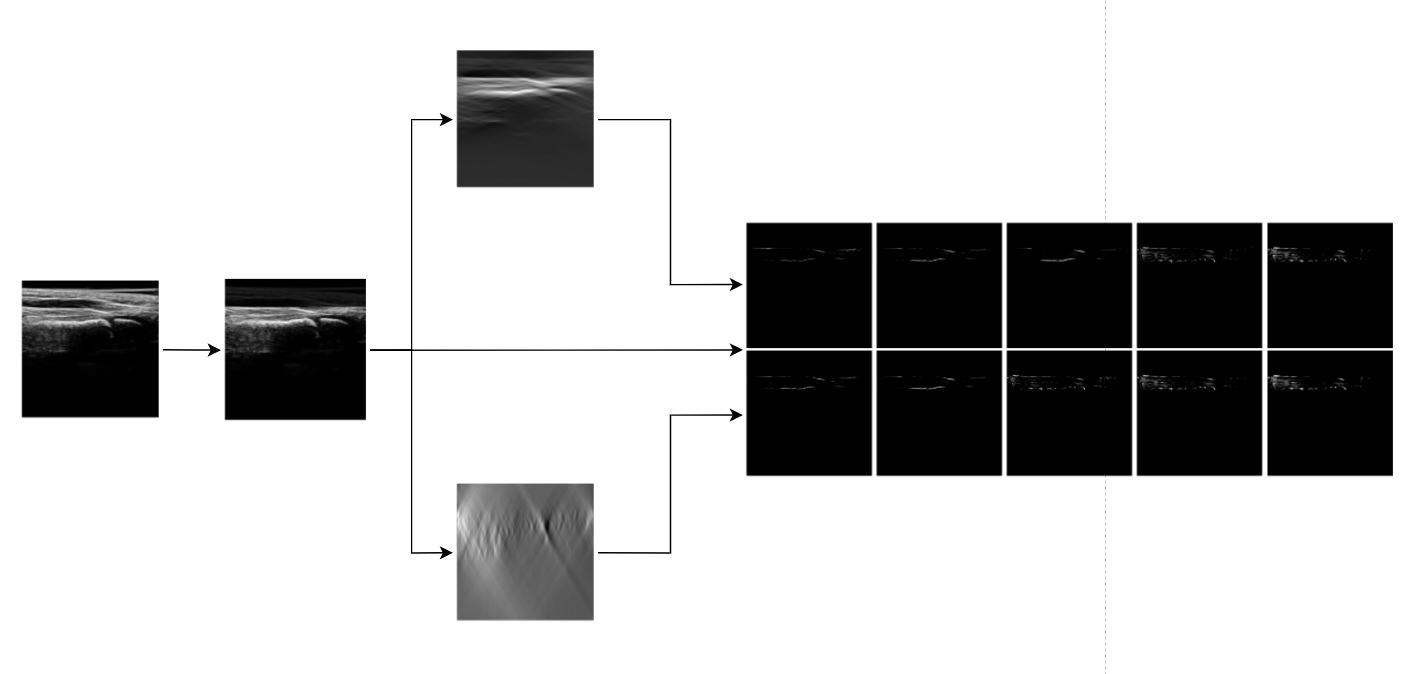}
\setlength{\belowcaptionskip}{-10pt}
\caption{DGA + RT-FPM preprocessing pipeline (refer Figure 1) of Wrist Ultrasound frame, note that DGA attenuates the upper section of the raw frame in this case}
\label{fig_WUS_preprocess}
\end{figure*}

\begin{figure*}[!h]
\centering
\includegraphics[width=\textwidth]{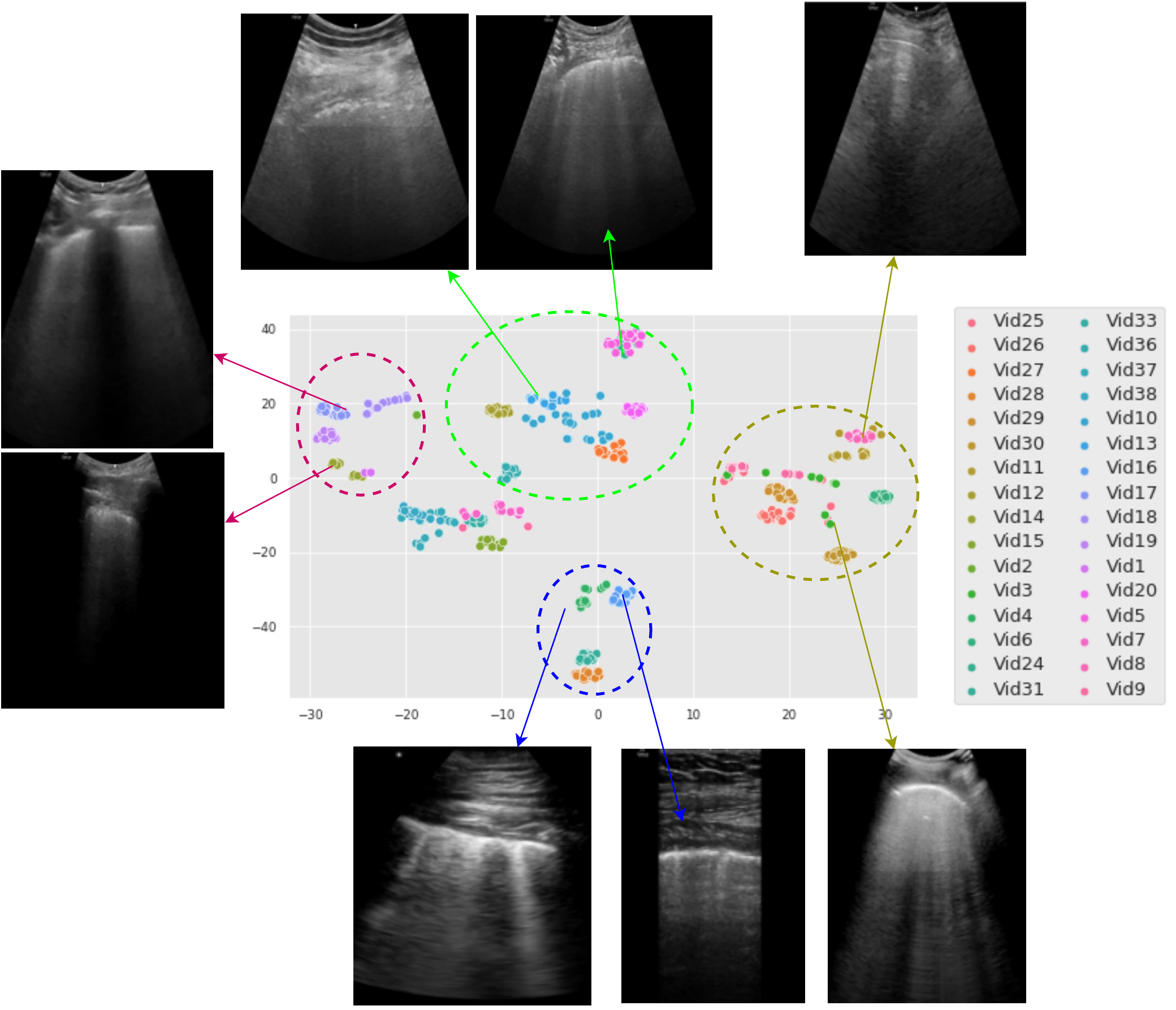}
\caption{ t-SNE plot on LUS data with  FF-CNN  embedding  of  Transporter  with  Attention  with frame sample to video instance labels. The plot clearly shows the clusters being formed based on frame similarity which shows the potential application of unsupervised clustering of video frames based on similarity}

\label{fig:VideoClass}
\end{figure*}